\documentclass[%
 reprint,
superscriptaddress,
 amsmath,amssymb,
prl
]{revtex4-1}

\usepackage{graphicx}
\usepackage{subfigure}
\usepackage{dcolumn}
\usepackage{bm}
\usepackage{color}

\def\nin{\noindent} 
\def\beq{\begin{equation}}
\def\eeq{\end{equation}}

\begin{document}


\title{Structural evolution of granular systems: Theory}

\author{Raphael Blumenfeld}%
 \email{rbb11@cam.ac.uk}
\affiliation{
Earth Science and Engineering, Imperial College London, London SW7 2AZ, UK\\
}%
\affiliation{
Cavendish Laboratory, Cambridge University, JJ Thomson Avenue, Cambridge CB3 0HE, UK\\
}%
\affiliation{
College of Science, NUDT, Changsha, Hunan, China
}%

\date{\today}

\begin{abstract}

A first-principles theory is developed for the general evolution of a key structural characteristic of planar granular systems - the cell order distribution. The dynamic equations are constructed and solved in closed form for a number of examples: dense systems undergoing progressive compaction; initial dilation of very dense systems; and the approach to steady state of general systems. 
It is shown that the convergence to steady state is exponential, except when contacts are only broken and no new contacts are made, in which case the approach is algebraic in time.
Where no closed form solutions are possible, illustrative numerical solutions of the evolution are shown. These show that the dynamics are sensitive to the cell event rates, which are process dependent.
The formalism can be extended to other structural characteristics, paving the way to a general theory of structural organisation of granular systems, parameterised by the contact event rates.

\end{abstract}

\pacs{45.70.Cc, 45.70.Vn, 81.05.Rm, 05.65.+b}

\keywords{Granular dynamics, structural evolution, cell order distribution}

\maketitle

\nin {\bf Introduction:} 
Understanding and modelling the self-organization of granular matter under external forces is essential to many natural phenomena and technological applications. Examples are: consolidation and failure of granular matter, packing of particulates, initiation of avalanches, flow of slurry and dense colloidal suspensions, to mention a few. 
This is arguably the most important problem in the field of granular science \cite{EdOa89a, Vogel_Roth_2003, Cheng_etal_1999, BaBl02, Aste_etal_2007, Meyer-etal2010}. 
During the evolution of dense granular systems, intergranular contacts are continually made and broken, the structure is constantly changing and so is the manner by which forces are transmitted through the medium. 

The structure of a granular system determines its stress transmission and affects a wide range of physical properties.
For example, void size distribution and connectivity determine the permeability to fluid flow, which is relevant to underground water, pollutant dispersion and oil extraction \cite{Vogel_Roth_2003}. Different structures also give rise to different solid-void surface distribution \cite{Cheng_etal_1999}, which is significant to catalysis, heat exchange between the solid and the void space, and to the functionality of fuel cell electrodes. Thus, predicting the structure that granular materials settle into is essential and the absence of a theory for the organisation dynamics of granular matter is a major obstacle to better modelling of a wide range of engineering and technological applications. 
This paper addresses this problem - a basic theory is developed for the evolution of the structure of two-dimensional (2D) systems of rigid grains. 

The structure is mainly defined by an inter-granular contact network, which, in 2D, delineates voids, or cells. A basic characteristic of a cell is its order, i.e. the number of grains that enclose it.  This paper focuses on the evolution of the cell order distribution (COD). The evolution equations are constructed and solved in closed form, under some assumptions, both for very dense systems and for systems approaching a limit steady state. Numerical solutions are also presented for more general cases. 
The formalism developed here can be extended to describe evolution of other characteristics, such as a cell volumes and local descriptors, such as quadrons \cite{BaBl02,BlEd03}, as outlined briefly in the concluding discussion. \\

\begin{figure}
\includegraphics[width=8cm]{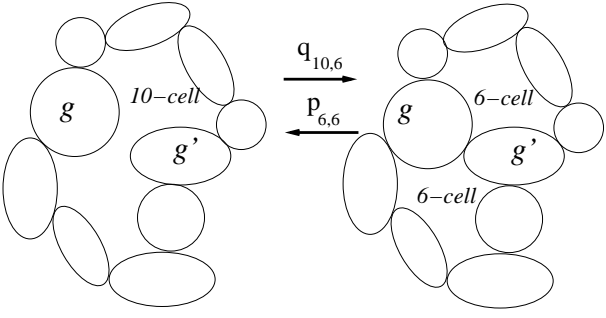}%
\caption{\label{Fig1} 
Making a contact between $g$ and $g'$, taking place at rate $p_{6,6}$, breaks a $10$-cell into two $6$-cells (left to right). Conversely, breaking this contact, at a rate $q_{10,6}$, merges two $6$-cells into a $10$-cell.}
\end{figure}

\nin {\bf The evolution equations:} 
The dynamics of the structural organisation are presumed to be slow and quasi-static, such that the integrity of the contact network is maintained and cells can be defined. 
As the structure evolves, every creation or breaking of an inter-granular contact, henceforth a contact event (CE), leads to splitting or merging of cells, respectively. 
Two neighbour cells always share a contact and, when this contact breaks, the cells merge. If the originals were cells of order $i$ and $j$, or $i$- and $j$-cells, for brevity, then this process generates a new $(i+j-2)$-cell (see e.g. Figure \ref{Fig1}). 
Conversely, when a new contact is made, a $k$-cell is `pinched' into an $i$-cell and a $j$-cell, such that $i+j=k+2$. 
The following analysis focuses on the evolution of the COD, which is central to the packing problem \cite{Toetal} and has been argued \cite{Fretal08} and shown \cite{MaBl14} to converge to a universal form when the intergranular friction is scaled away. To develop a theory for this distribution, the following need to be defined: \\
$p_{i,k-i+2}$ - the rate at which an $i$-cell and a $(k-i+2)$-cell combine to make a $k$-cell; \\
$q_{k,i}$ - the rate at which a $k$-cell breaks into an $i$-cell and a $(k-i+2)$-cell; \\
$r_{i,k-i+4}$ - the rate at which an $i$-cell and a $(k-i+4)$-cell combine to make a $k$-cell with exactly one rattler within; \\
$s_{k,i}$ - the rate at which a $k$-cell, containing a rattler, breaks into an $i$-cell and a $(k-i+4)$-cell; \\
$\rho_k$ - the number density of $k$-cells.\\
The COD evolves via the following basic CEs:\\
(i) $k$-cell creation by joining of two neighbouring cells of orders $3\le i<k$ and $3\le k-i+2<k$ at a rate $p_{i,k-i+2}$;\\
(ii) $k$-cell creation by pinching of an $i$-cell into two cells of orders $k<i$ and $i-k+2<i$ at a rate $q_{i,k}$;\\
(iii) $k$-cell pinched into two cells of orders $i<k$ and $k-i+2<k$ at a rate $q_{k,i}$;\\ 
(iv) $k$-cell elimination due to combining with a neighbouring $(i-k+2)$-cell to make an $i$-cell ($i>k$) at a rate $p_{k,i-k+2}$.\\
Each of these processes has an equivalent when the combined cell contains a rattler, in which case the rates $p_{j,k-j+2}$ and $q_{k,j}$ are replaced, respectively, by the rates $r_{j,k-j+4}$ and $s_{k,j}$. 
It is assumed in the following that very large cells are very rare, allowing us to limit the discussion to dynamics of cells containing at most one rattler. 
The evolution equations are then
\begin{eqnarray}
\dot{\rho_k} & = &\frac{1}{2}\big\{ \sum_{i=3}^{k-1} \left[ \rho_i\rho_{k-i+2}p_{i,k-i+2} - \rho_k q_{k,i}\right]\left(1 + \delta_{i,k-i+2}\right) + \nonumber \\
& + & \left[\rho_i\rho_{k-i+4}r_{i,k-i+4} - \rho_k s_{k,i}\right]\left(1 + \delta_{i,k-i+4}\right)\big\} + \nonumber \\
& + & \sum_{i=k+1}^{\infty} \big\{\left(\rho_i q_{i,k} - \rho_k\rho_{i-k+2}p_{k,i-k+2}\right)\left(1+\delta_{i,2k-2}\right)  + \nonumber \\
& + & \left(\rho_i s_{i,k} - \rho_k\rho_{i-k+4}r_{k,i-k+4}\right)\left(1+\delta_{i,2k-4}\right) \big\} 
\label{eqmaster1}
\end{eqnarray}
The terms $\left(1 + \delta_{i,k-i+2}\right)/2$ and $\left(1 + \delta_{i,k-i+4}\right)/2$ assure correct counting and the terms $\left(1 + \delta_{i,2k-2}\right)$ and $\left(1 + \delta_{i,2k-4}\right)$ describe, respectively, generation and disintegration of two $k$-cells from and into two cells of equal order. 
Note that $\rho_k$ has units of inverse volume and, therefore, that $p_{i,j}$ and $r_{i,j}$ have units of (volume/time) and $q_{i,k}$ and $s_{ij}$ of inverse time.  
For clarity of the following analysis, I ignore rattlers and set $r_{i,j}=s_{ij}=0$. 
Including rattlers, which is significant for modelling many realistic applications, is straightforward, but it would not add insight beyond the results presented below. \\

\nin {\bf Exact solutions:} 
To illustrate the usefulness of the evolution equation (\ref{eqmaster1}), let us first consider a simple case: the evolution of a granular system containing only 3- and 4-cells. This could be a model for  the dense end stages of a compression process, when the mean number of contacts per grain is $4\leq z\leq 6$ and all the higher order cells have split. 
For a grain size distribution, which is not too broad, such small cells cannot include rattlers and equations (\ref{eqmaster1}) reduce to
\begin{eqnarray}
\dot{\rho_3} & = & 2 \left(q_{4,3}\rho_4  - p_{3,3}\rho_3^2\right) \label{eqmaster4a} \\
\dot{\rho_4} & = & p_{3,3}\rho_3^2 - q_{4,3}\rho_4 = -\frac{1}{2} \dot{\rho_3}  
\label{eqmaster4b}
\end{eqnarray}
The first thing to note is that these equations satisfy a conservation law; $2\rho_4 + \rho_3 = C_0$, where $C_0$ is determined by initial conditions. 
This is, in fact, a general feature of eqs. (\ref{eqmaster1}) - a system containing up to $N$-cells satisfies 
\beq
\sum_{k=3}^N (k-2)\rho_k = C_0
\label{GeneralConservation}
\eeq
where $C_0$ is a constant determined by the initial state. This constant has a physical meaning: it is the number of 3-cells that would be generated if all the cells broke eventually into smaller ones. Thus, $C_0$ corresponds to the state of maximal compaction, in which the mean coordination number per grain is six, up to negligible boundary corrections.
Without loss of generality, it is possible to normalise the densities by this constant, $\rho_k\to \rho_k/C_0$,
giving $\sum_{k=3}^N (k-2)\rho_k = 1$. In these units, the densities are bounded, $0\leq \rho_k\leq 1/(k-2)$. 

\begin{figure}
\includegraphics[width=4.2cm]{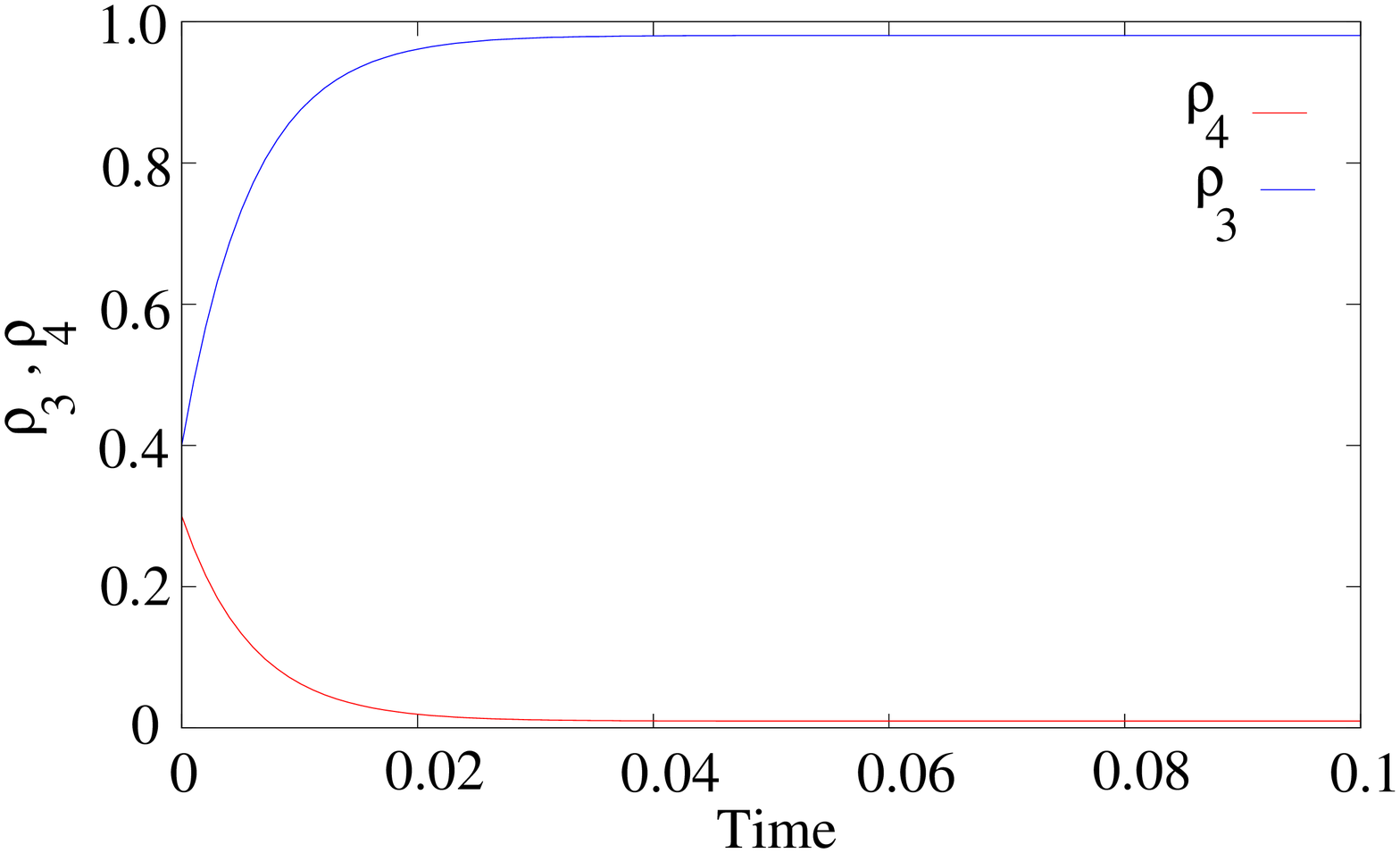}%
\includegraphics[width=4.2cm]{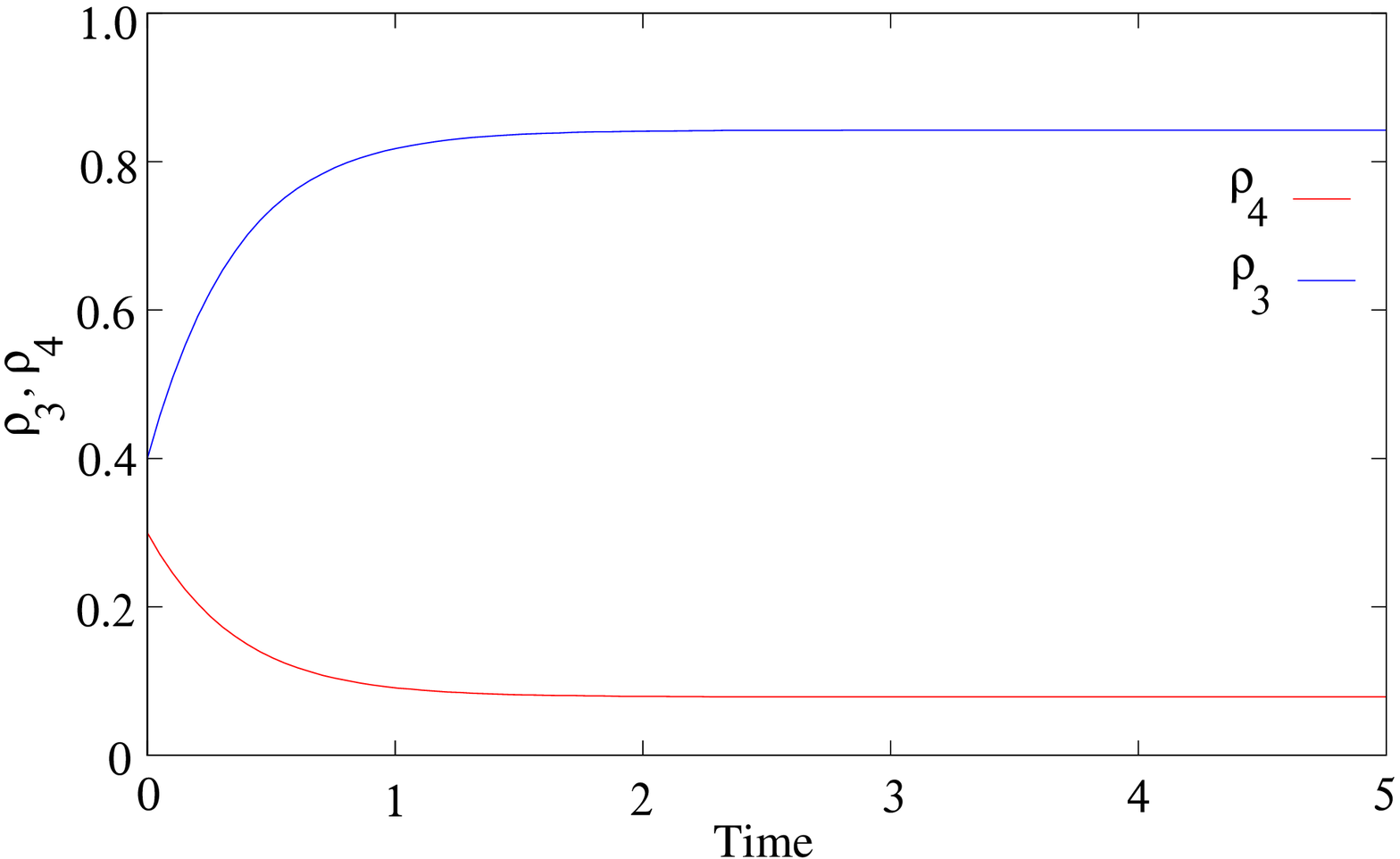}%
\caption{\label{Fig2} 
Examples of the evolution of dense $\rho_3-\rho_4$ systems. \\
Left: $p_{3,3}=0.01;\ q_{4,3}=0.99;\ \rho_3(0)=0.4;\ \rho_4(0)=0.3$; \\
right: $p_{3,3}=0.1;\ q_{4,3}=0.9;\ \rho_3(0)=0.4;\ \rho_4(0)=0.3$. \\
Note the different convergence rates.}
\end{figure}

Eqs. (\ref{eqmaster4a})-(\ref{eqmaster4b}) can be solved exactly under the assumption that the CE rates, $p_{3,3}$ and $q_{4,3}$, are time-independent. 
Using the conservation law to eliminate $\rho_3$ between the equations and rearranging gives
\beq
\dot{\rho_4} = 4p_{3,3}\left[\left(\rho_4-\alpha\right)^2 - \beta^2\right]
\label{rho4EvolutionEq}
\eeq
where $\alpha=\frac{1}{2}\left(\frac{q_{4,3}}{4p_{3,3}}+C_0\right)$ and $\beta^2=\left(\frac{q_{4,3}}{8p_{3,3}}\right)^2+\frac{q_{4,3}C_0}{8p_{3,3}}$.
This equation can be solved in closed form:
\beq
\rho_4(t) = \alpha - \beta + \frac{2\beta}{1 + \gamma e^{8 p_{3,3}\beta t}}
\label{rho4EvolutionEq1}
\eeq
where 
\beq
\gamma = \frac{\beta - \alpha + \rho_4(0)}{\beta + \alpha - \rho_4(0)}
\label{gamma}
\eeq
Note that the negative and positive values of $\beta$ yield the same solution. 
The time dependence of $\rho_3(t)$ is obtained by using (\ref{rho4EvolutionEq1}) and the conservation law. 
A plot of $\rho_3(t)$ and $\rho_4(t)$ for two different initial conditions and CE rates is shown in Figure \ref{Fig2}.
With time, the evolution converges to a steady state exponentially, a generic behaviour that will be discussed below.
For example, a particular model of progressive compaction would be by not letting 3-cells merge at all, i.e. $p_{3,3}=0$. Solution (\ref{rho4EvolutionEq1}) then reduces to
\beq
\rho_4(t) = \rho_4(0) e^{-q_{4,3}t}
\label{rho4EvolutionEq2}
\eeq

However, not all approaches to steady state are exponential. For example, consider the dilation of a very dense system under shear. As mentioned, the densest state of planar systems consist of only 3- and 4-cells; the denser the system the larger the fraction of 3-cells. Under applied shear, such systems will dilate, which means that contacts will be broken and 4-cells will be generated. Under such conditions, splitting of 4-cells back into 3-cells is rare. An idealised model of this process would be by setting $q_{4,3}=0$. Then, the solution of eqs. (\ref{eqmaster4a})-(\ref{eqmaster4b}) is
\beq
\rho_3(t) = \frac{\rho_3(0)}{1 + 2p_{3,3}\rho_3(0)t} = 1 - 2\rho_4(t)
\label{rho3EvolutionEq2}
\eeq
and the final state consists of only 4-cells. We see that, unlike the generic case, this steady state is approached algebraically.

\begin{figure}
\includegraphics[width=4.25cm]{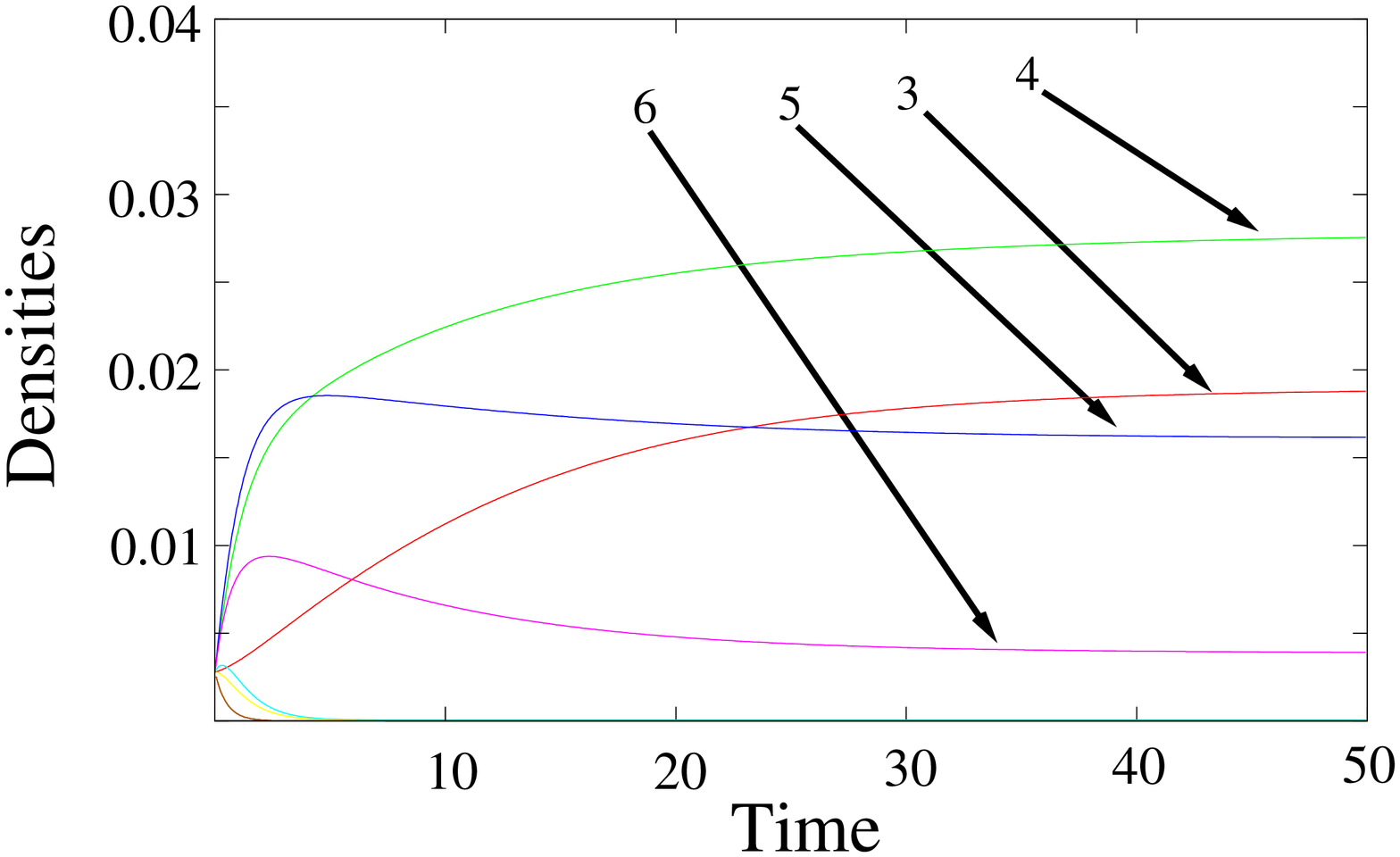}
\includegraphics[width=4.25cm]{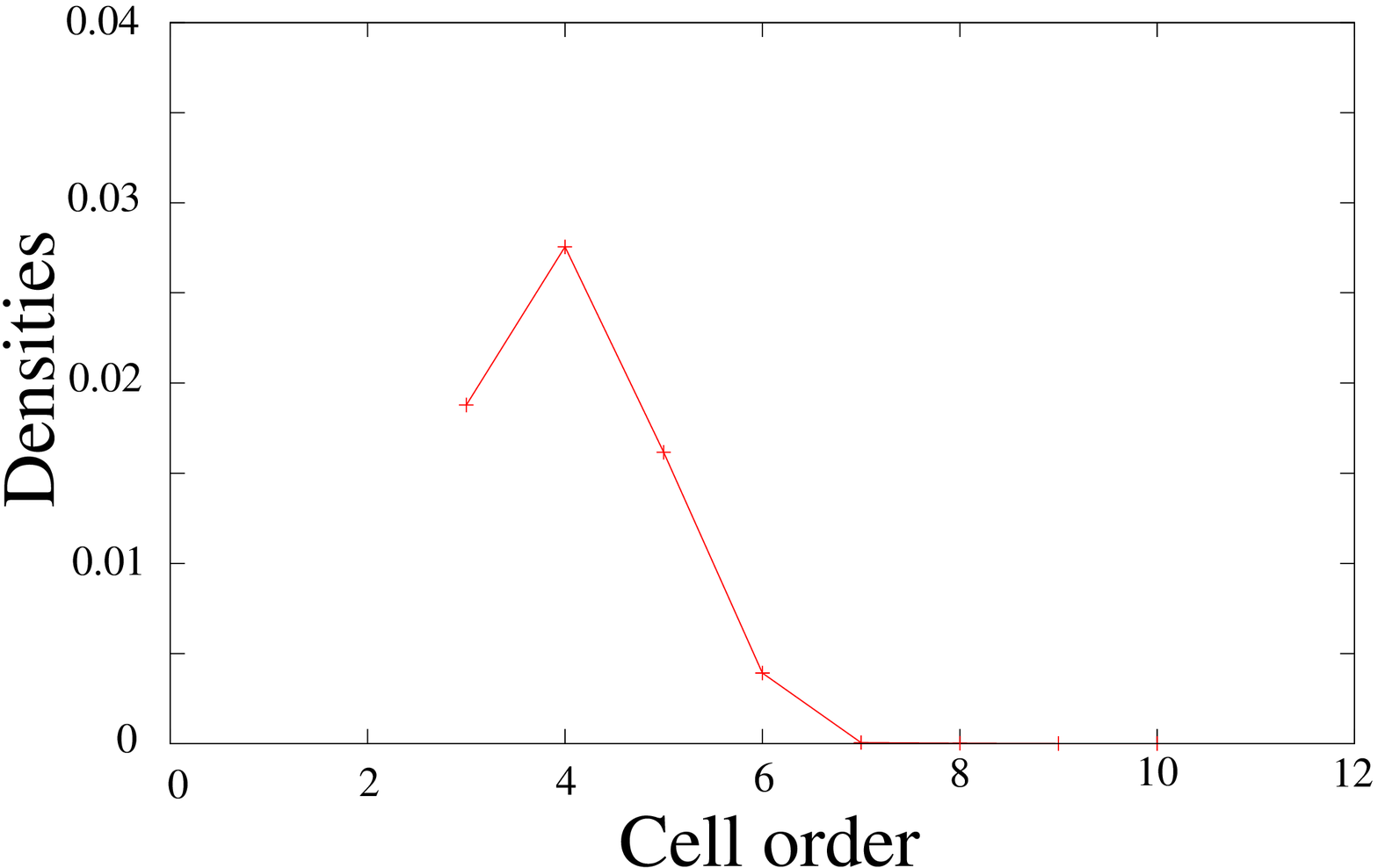}%
\caption{\label{Fig3} 
The evolution of $\rho_3-\rho_{10}$ with $p_{i,j}$ and $\ q_{i,j}$ taken from parallel-plate shear simulations. Note the agreement with figure 12 of \cite{MaBl14}.}
\end{figure}

A closed form solution for the evolution of systems containing 3-, 4- and 5-cells has also been derived. However, it cumbersomely involves a solution of a cubic equation and does not provide further significant insight. \\
Somewhat more informative is to examine systems with many cell types. Figure \ref{Fig3} shows a numerical solution to the evolution of a system containing $\rho_3-\rho_{10}$. It is found that the details of the resultant COD depend on the chosen CE rates, suggesting that different processes with different rates are likely to lead to widely different COD's.  

\nin {\bf The approach to steady state:} 
The exponential approach of the COD to a steady state is due to the linear terms on the right hand side of eqs. (\ref{eqmaster1}), involving cell splitting. 
At the steady state, the left hand side of (\ref{eqmaster1}) vanishes and the densities converge to $\rho_k^{s}$. 
A particular steady-state solution is when each term in the sum in (\ref{eqmaster1}) vanishes, in which case the ratio of merge and break steady state rates satisfy
\beq
\frac{q_{k,i}^{s}}{p_{i,k-i+2}^{s}} = \frac{\rho_i^{s}\rho_{k-i+2}^{s}}{\rho_k^{s}}
\label{ssrelation}
\eeq
Before reaching the steady state, the densities are $\rho_k = \rho_k^{s}(1 + f_k(t))$, with $f_k(t)\ll 1$.  Expanding (\ref{eqmaster1}) to first order in the $f_k$'s and using (\ref{ssrelation}), yields a linear set of equations
\beq
\dot{f_k} = A_{kn}f_n \qquad {\rm or} \qquad  \dot{F}=A\cdot F
\label{LinEqs}
\eeq
with 
\begin{widetext}
\beq
A_{kn} = 
 \begin{cases}
  \frac{1} {2}p_{n,k-n+2}\rho^s_{n,k-n+2}  & n\leq k-1 \ \ \ ;\ \ \ n\neq n-k+2 \\
  -\frac{1}{2}\sum_{i=3}^{k-1}q_{k,i}\left(1+\delta_{k,2i-2}\right) - \sum_{i=k+1}^{N}p_{k,i-k+2}\rho^s_{k,i-k+2} \left(1+\delta_{k,2i-2}\right)  & n=k \\
  q_{n,k} - p_{k,n}\rho^{s}_k & n\geq k+1
\end{cases}
\label{ATerms}
\eeq
\end{widetext}
where we have set the largest possible cell order to $N$ and dropped the superscript $s$ from the CE rates, for brevity. 
Using the conservation law (\ref{GeneralConservation}) reduces these to a linear set of $N-1$ equations. 
For example, for the 3-4 system discussed above  
\begin{equation}
A=
\left( \begin{array}{cc} -4p_{3,3}\rho^s_3 & 2q_{4,3}\frac{\rho^s_4}{\rho^s_3}\cr \frac{2p_{3,3}(\rho^s_3)^2}{\rho^s_4} & -q_{4,3} \end{array} \right)
\label{AMatrix}
\end{equation}
which, with the conservation law, can be reduced to only one equation for, say, $\dot{f}_3$
The reduced set (\ref{LinEqs}) can be diagonalised, $\Lambda = D\cdot A\cdot D^{-1}$, $\Lambda_{ik} = \delta_{ik}\lambda_i$, with $\lambda_i$ the eigenvalues of $A$. The exact solution for $F$ is
\beq
f_k = \sum_i D_{ki} u_i(t=0) e^{\lambda_i t}
\label{SEvolution}
\eeq
where $u_i(t=0) = \sum_j \left(D^{-1}\right)_{ij} f_j(t=0)$.
For the 3-4 system, the solution is 
\beq
f_3 \sim e^{-\nu_1 t} \quad ;\quad f_4 = 1 - 2f_3
\label{SEvolution34}
\eeq
with $\nu_1 = \left(4p_{3,3} + q_{4,3}\right)\rho^s_3$. 
In  all the numerical solution generated for systems of up to cells of order 10, only real and negative eigenvalues were found, $\lambda_i\equiv -\nu_i \leq 0$. That this is the general case is difficult to prove directly from the form of (\ref{ATerms}). Complex eigenvalues could lead to steady state oscillations, while any positive eigenvalue would increase the density of a particular $k$-cell at the expense of all others until it reaches the maximum allowed by the conservation law, in which case the distribution converges to a delta-function. 
For the observed negative and real eigenvalues in the numerical solutions, the rate of convergence to the steady state is dictated by the smallest $\nu$-value: $\nu_{m}\equiv {\rm min}\left\{ \nu_i\right\}$,
\beq
f_j(t\to\infty) = D_{jm} e^{-\nu_m t} \quad {\rm (no\ summation\ over\ {\it m})}
\label{AsymptoticForm}
\eeq
which can be observed in Figure \ref{Fig3}. \\

\nin {\bf Conclusions and discussion:} 
To conclude, a formalism has been developed to describe the evolution of a significant structural characteristic of planar granular assemblies, the cell order distribution.
A basic et of equations has been developed and solved for the COD in closed form for several special cases and for general systems approaching a steady state. The closed form solution for the 3- and 4-cells system could be applicable to granular systems towards the end of a compression process, suggesting a way to test these results experimentally. \\
A numerical solution for systems up to 10-cells, has also been presented, which agrees with the distributions obtained in the literature for polydisperse disc systems near marginally rigid \cite{MaBl14}. In all these cases the details of the evolution and the final steady states depend sensitively on the CE rates. \\
It has been shown that the approach to the steady state is generically exponential due to the terms involving generation of intergranular contacts. In the absence of these terms, i.e. when contacts only break and small cells merge, the solution has been shown to be algebraic, at least for the 3-4 system. Realistically, such dynamics could model the initial dilation of shear bands, again suggesting a way to test these results. 

The CE rates have been chosen to be time-independent mainly to illustrate the usefulness of this, proof-of-principle, formalism. In reality, the local probability of a CE between grains may depend on the forces acting on them by other grains, as well as on the orders and shapes of the cells that they belong to. Taking these into consideration would lead to a nonlinear theory and it is the direction to be explored next. Extending the formalism to include dependence of CEs on local forces is significant because it paves the way to a fundamental self-consistent model of the coupled evolution of the structure and the force distribution\cite{ForceStructConsistency}.

The formalism developed here can be applied widely to describe the evolution of other structural characteristics during the organisation of granular mater. For example, it can be used to describe the statistics of the elementary volume elements - the quadrons \cite{BlEd03,BlEd06}. Making a contact corresponds to two quadrons splitting into four and visa versa when a contact is broken. 
The evolution eqs. (\ref{eqmaster1}) could then describe, after a straightforward modification, the total and conditional quadron volume distributions, which have been found recently to exhibit universal behaviour \cite{MaBl14}. 


\end{document}